\documentclass[12pt,a4paper,fleqn,notitlepage,onecolumn]{article}

\usepackage[english]{babel}
\usepackage[ansinew]{inputenc}
\usepackage{amssymb,amsmath}
\usepackage{graphicx} % Package zum Einfuegen von Grafiken (u.a. pdf-Format)
\usepackage{fancyhdr} %fuer fancy header
\usepackage{natbib} % fuer automatische Verwaltung der Referenzen
\usepackage{wrapfig}
\usepackage{geometry}
\usepackage{ulem}
\usepackage{color}
\newtheorem{example}{Example}

\newcommand{\prob}{\textrm{prob}}

\newcommand{\diag}{\textrm{diag}}

\newcommand{\mat}{\boldsymbol}

\title{Dos and don'ts of reduced chi-squared}

\author{Ren\'e Andrae${}^1$, Tim Schulze-Hartung${}^1$ \& Peter Melchior${}^2$\\
\footnotesize{${}^1$ Max-Planck-Institut f\"ur Astronomie, K\"onigstuhl 17, 69117 Heidelberg, Germany}\\
\footnotesize{${}^2$ Institut f\"ur Theoretische Astrophysik, ZAH, Albert-Ueberle-Str. 2, 69120 Heidelberg, Germany}\\
\footnotesize{e-mail: andrae@mpia-hd.mpg.de}}
\date{}

\begin{document}

\maketitle

\begin{center}
\begin{minipage}{15cm}
\small{Reduced chi-squared is a very popular method for model assessment, model comparison, convergence diagnostic, and error estimation in astronomy. In this manuscript, we discuss the pitfalls involved in using reduced chi-squared. There are two independent problems: (a) The number of degrees of freedom can only be estimated for linear models. Concerning nonlinear models, the number of degrees of freedom is unknown, i.e., it is not possible to compute the value of reduced chi-squared. (b) Due to random noise in the data, also the value of reduced chi-squared itself is subject to noise, i.e., the value is uncertain. This uncertainty impairs the usefulness of reduced chi-squared for differentiating between models or assessing convergence of a minimisation procedure. The impact of noise on the value of reduced chi-squared is surprisingly large, in particular for small data sets, which are very common in astrophysical problems. We conclude that reduced chi-squared can only be used with due caution for linear models, whereas it must not be used for nonlinear models at all. Finally, we recommend more sophisticated and reliable methods, which are also applicable to nonlinear models.}
\end{minipage}
\end{center}

\section{Introduction}

When fitting a model $f$ with parameters $\vec\theta$ to $N$ data values $y_n$, measured with (uncorrelated) Gaussian errors $\sigma_n$ at positions $\vec x_n$, one needs to minimise
\begin{equation}\label{eq:def:chi2}
\chi^2 = \sum_{n=1}^N \left(\frac{y_n - f(\vec x_n;\vec\theta)}{\sigma_n}\right)^2 \;\textrm{.}
\end{equation}
This is equivalent to maximising the so-called ``likelihood function''. If the data's measurement errors are not Gaussian, $\chi^2$ should not be used because it is not the maximum-likelihood estimator. For the rest of this manuscript, we shall therefore assume that the data's errors are Gaussian. If $K$ denotes the number of degrees of freedom, reduced $\chi^2$ is then defined by
\begin{equation}
\chi^2_\textrm{red} = \frac{\chi^2}{K} \;\textrm{.}
\end{equation}
$\chi^2_\textrm{red}$ is a quantity widely used in astronomy. It is essentially used for the following purposes:
\begin{enumerate}
\item Single-model assessment: If a model is fitted to data and the resulting $\chi^2_\textrm{red}$ is larger than one, it is considered a ``bad'' fit, whereas if $\chi^2_\textrm{red}<1$, it is considered an overfit.
\item Model comparison: Given data and a set of different models, we ask the question which model fits the data best. Typically, each model is fit to the data and their values of $\chi^2_\textrm{red}$ are compared. The winning model is that one whose value of $\chi^2_\textrm{red}$ is closest to one.
\item Convergence diagnostic: A fit is typically an iterative process which has to be stopped when converged. Convergence is sometimes diagnosed by monitoring how the value of $\chi^2_\textrm{red}$ evolves during the iteration and the fit is stopped as soon as $\chi^2_\textrm{red}$ reaches a value sufficiently close to one. Sometimes it is claimed then, that ``the fit has reached noise level''.
\item Error estimation: One fits a certain model to given data by minimising $\chi^2$ and then rescales the data's errors such that the value of $\chi^2_\textrm{red}$ is exactly equal to one. From this one then computes the errors of the model parameters. (It has already been discussed by \citet{Andrae2010d} that this method is incorrect, so we will not consider it any further here.)
\end{enumerate}
In all these cases, $\chi^2_\textrm{red}$ excels in simplicity, since all one needs to do is divide the value of $\chi^2$ by the number of degrees of freedom and compare the resulting value of $\chi^2_\textrm{red}$ to one.

In this manuscript, we want to investigate the conditions under which the aforementioned applications are meaningful -- at least the first three. In particular, we discuss the pitfalls that may severly limit the credibility of these applications. We explain the two major problems that typically arise in using $\chi^2_\textrm{red}$ in practice: First, we dicuss the issue of estimating the number of degrees of freedom in Sect. \ref{sect:dofs}. Second, we explain how the uncertainty in the value of $\chi^2$ may affect the above applications in Sect. \ref{sect:noise_on_chi2}. Section \ref{sect:alternatives} is then dedicated to explain more reliable methods rather than $\chi^2_\textrm{red}$. We conclude in Sect. \ref{sect:conclusions}.

\section{Degrees of freedom\label{sect:dofs}}

Given the definition of $\chi^2_\textrm{red}$, it is evidently necessary to know the number of degrees of freedom of the model. For $N$ data points and $P$ fit parameters, a na\"ive guess is that the number of degrees of freedom is $N-P$. However, in this section, we explain why this is \textit{not} true in general. We begin with a definition of ``degrees of freedom'' and then split this discussion into three parts: First, we discuss only linear models. Second, we discuss linear models with priors. Third, we discuss nonlinear models. Finally, we discuss whether linearisation may help in the case of nonlinear models. 

\subsection{Definition\label{sect:def_dofs}}

For a given parameter estimate, e.g., a model fitted to data, the degrees of freedom are the number of independent pieces of information that were used. The concept of ``degrees of freedom'' can be defined in different ways. Here, we give a general and simple definition. In the next section, we give a more technical definition that only applies to linear models.

Let us suppose that we are given $N$ measurements $y_n$ and a model with $P$ free parameters $\theta_1,\theta_2,\ldots,\theta_P$. The best-fitting parameter values are found by minimising $\chi^2$. This means we impose $P$ constraints of the type
\begin{equation}\label{eq:dof_constraints}
\frac{\partial\chi^2}{\partial\theta_p}=0 \qquad\qquad \forall p=1,2,\ldots,P
\end{equation}
onto our $N$-dimensional system. Hence, the number of degrees of freedom is $K=N-P$. At first glance, this appears to be a concise and infallible definition. However, we shall see that this is not the case.

\subsection{Linear models without priors\label{sect:dofs_linear}}

A linear model is a model, where \textit{all} fit parameters are linear. This means it is a linear superposition of a set of basis functions,
\begin{equation}
f(\vec x,\vec\theta) = \theta_1 B_1(\vec x)+\theta_2 B_2(\vec x)+\ldots+\theta_P B_P(\vec x) = \sum_{p=1}^P \theta_p B_p(\vec x)\;\textrm{,}
\end{equation}
where the coefficients $\theta_p$ are the fit parameters and the $B_p(\vec x)$ are some (potentially nonlinear) functions of the position $\vec x$. A typical example is a polynomial fit, where $B_p(x)=x^p$. Inserting such a linear model into $\chi^2$ causes $\chi^2$ to be a quadratic function of the fit parameters, i.e., the first derivatives -- our constraints from Eq.\ (\ref{eq:dof_constraints}) -- form a set of linear equations that can be solved analytically under certain assumptions.

A natural approach to solve such sets of linear equations is to employ linear algebra. This will lead us to a quantitative definition of the number of degrees of freedom for linear models. Let us introduce the following quantities:
\begin{itemize}
\item $\vec y=(y_1,y_2,\ldots,y_N)^T$ is the $N$-dimensional vector of measurements $y_n$.
\item $\vec\theta=(\theta_1,\theta_2,\ldots,\theta_P)^T$ is the $P$-dimensional vector of linear model parameters $\theta_p$.
\item $\mat\Sigma$ is the $N\times N$ covariance matrix of the measurements, which is diagonal in the case of Eq.\ (\ref{eq:def:chi2}), i.e., $\mat\Sigma=\diag(\sigma_1^2,\sigma_2^2,\ldots,\sigma_N^2)$.
\item $\mat X$ is the so-called design matrix which has format $N\times P$. Its elements are given by $X_{np}=B_p(\vec x_n)$, i.e., the $p$-th basis function evaluated at the $n$-th measurement point.
\end{itemize}
Given these definitions, we can now rewrite Eq.\ (\ref{eq:def:chi2}) in matrix notation,
\begin{equation}\label{eq:only4linear}
\chi^2 = (\vec y - \mat X\cdot\vec\theta)^T\cdot\mat\Sigma^{-1}\cdot(\vec y - \mat X\cdot\vec\theta) \;\textrm{.}
\end{equation}
Minimising $\chi^2$ by solving the constraints of Eq.\ (\ref{eq:dof_constraints}) then yields the analytic solution
\begin{equation}
\hat{\vec\theta} = (\mat X^T\cdot\mat\Sigma^{-1}\cdot\mat X)^{-1}\cdot\mat X^T\cdot\mat\Sigma^{-1}\cdot\vec y \;\textrm{,}
\end{equation}
where the hat in $\hat{\vec\theta}$ accounts for the fact that this is only an \textit{estimator} for $\vec\theta$, but not the true $\vec\theta$ itself. We then obtain the prediction $\hat{\vec y}$ of the measurements $\vec y$ by,
\begin{equation}\label{eq_y_hat}
\hat{\vec y} = \mat X\cdot\hat{\vec\theta} = \mat X\cdot(\mat X^T\cdot\mat\Sigma^{-1}\cdot\mat X)^{-1}\cdot\mat X^T\cdot\mat\Sigma^{-1}\cdot\vec y = \mat H\cdot\vec y \;\textrm{,}
\end{equation}
where we have introduced the $N\times N$ matrix $\mat H$, which is sometimes called ``hat matrix'', because it translates the data $\vec y$ into a model prediction $\hat{\vec y}$. The number of \textit{effective} model parameters is then given by the trace of $\mat H$ \citep[e.g.][]{Ye1998,Hastie2009},
\begin{equation}\label{eq:P_effective}
P_\textrm{eff} = \textrm{tr}(\mat H) = \sum_{n=1}^N H_{nn} = \textrm{rank}(\mat X) \;\textrm{,}
\end{equation}
which also equals the rank of the design matrix $\mat X$.\footnote{The rank of $\mat X$ equals the number of linearly independent column vectors of $\mat X$.} Obviously, $P_\textrm{eff}\leq P$, where the equality holds if and only if the design matrix $\mat X$ has full rank. Consequently, for linear models the number of degrees of freedom is
\begin{equation}
K=N-P_\textrm{eff}\geq N-P \;\textrm{.}
\end{equation}

The standard claim is that a linear model with $P$ parameters removes $P$ degrees of freedom when fitted to $N$ data points, such that the remaining number of degrees of freedom is $K=N-P$. Is this correct? No, not necessarily so. The problem is in the required linear independence of the $P$ basis functions. We can also say that the $P$ constraints given by Eq.\ (\ref{eq:dof_constraints}) are not automatically independent of each other. Let us consider a trivial example, where the basis functions are clearly not linearly independent:
\begin{example}\label{ex_1}
The linear model $f(\vec x,\vec\theta) = \theta_1+\theta_2$ is composed of two constants, $\theta_1$ and $\theta_2$, i.e., $B_1(\vec x)=B_2(\vec x)=1$. Obviously, this two-parameter linear model cannot fit two arbitrary data points and its number of degrees of freedom is not given by $N-2$ but $N-1$, because the design matrix $\mat X$ only has rank 1, not rank 2. In simple words, the two constraints $\frac{\partial\chi^2}{\partial\theta_1}=0$ and $\frac{\partial\chi^2}{\partial\theta_2}=0$ are not independent of each other.
\end{example}
From this discussion we have to draw the conclusion that for a linear model the number of degrees of freedom is given by $N-P$ \textit{if and only if} the basis functions are linearly independent \textit{for the given data positions $\vec x_n$}, which means that the design matrix $\mat X$ has full rank. In practice, this condition is \textit{usually} satisfied, but not always. In the more general case, the true number of degrees of freedom for a linear model may be anywhere between $N-P$ and $N-1$.

\begin{wrapfigure}{right}{8cm}
\includegraphics[width=8cm]{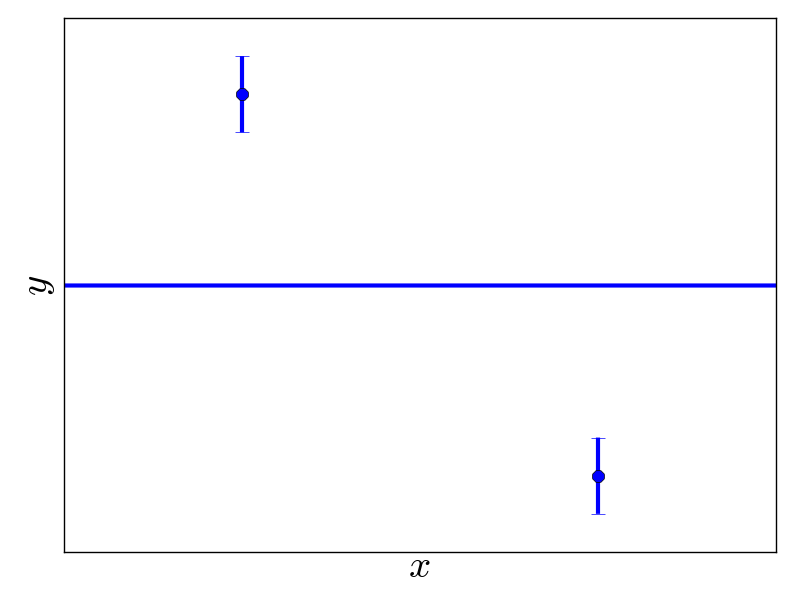}
\caption{Example of a two-parameter model, $f(x)=a_0+a_1 x$, that is incapable of fitting two data points perfectly because it involves a prior $a_1\geq 0$. We show the optimal fit for the given data. The model is discussed in Example \ref{ex_2}.}
\label{fig:linear_model_with_prior}
\end{wrapfigure}

\subsection{Linear models with priors}

Priors are commonly used to restrict the possible values of fit parameters. In practice, priors are usually motivated by physical arguments, e.g., a fit parameter corresponding to the mass of an object must not be negative. Let us consider a very simple example of a prior:
\begin{example}\label{ex_2}
A linear model $f(x,a_0,a_1)=a_0+a_1 x$ is given. The value of parameter $a_0$ is not restricted, but $a_1$ must not be negative. Figure \ref{fig:linear_model_with_prior} demonstrates that this two-parameter model is incapable of sensibly fitting two arbitrary data points because of this prior.
\end{example}
Obviously, priors reduce the flexibility of a model, which is actually what they are designed to do. Consequently, they also affect the number of degrees of freedom. In this case, the prior was a step function (zero for $a_1<0$ and one otherwise), i.e., it was highly nonlinear. Consequently, although $f(x,a_0,a_1)=a_0+a_1 x$ is itself a linear function of all fit parameters, the overall model including the prior is not linear anymore. This leads us directly to the issue of nonlinear models.

\subsection{Nonlinear models\label{sect:dofs_nonlinear}}

We have seen that estimating the number of degrees of freedom is possible in the case of linear models with the help of Eq.\ (\ref{eq:P_effective}). However, for a nonlinear model, we cannot rewrite $\chi^2$ like in Eq.\ (\ref{eq:only4linear}), because a nonlinear model cannot be written as $\mat X\cdot\vec\theta$. Therefore, $\mat H$ does not exist and we cannot use Eq.\ (\ref{eq:P_effective}) for estimating the number of degrees of freedom. \citet{Ye1998} introduces the concept of ``generalised degrees of freedom'', but concludes that it is infeasible in practice. We now consider two examples in order to get an impression why the concept of degrees of freedom is difficult to grasp for nonlinear models:
\begin{example}\label{ex_3}
Let us consider the following model $f(x)$, having three free parameters $A$, $B$, $C$,
\begin{equation}\label{eq:circular_orbit}
f(x) = A\cos(B x+C) \;\textrm{.}
\end{equation}
If we are given a set of $N$ measurement $(x_n,y_n,\sigma_n)$ such that no two data points have identical $x_n$, then the model $f(x)$ is capable of fitting \emph{any} such data set \emph{perfectly}. The way this works is by increasing the ``frequency'' $B$ such that $f(x)$ can change on arbitrarily short scales.\footnote{In practice, there is of course a prior forbidding unphysically large frequencies. But there is no such restriction in this thought experiment.} As $f(x)$ provides a perfect fit in this case, $\chi^2$ is equal to zero for all possible noise realisations of the data. Evidently, this three-parameter model has \emph{infinite} flexibility (if there are no priors) and $K=N-P$ is a poor estimate of the number of degrees of freedom, which actually is $K=0$.
\end{example}
\begin{example}\label{ex_4}
Let us modify the model of Example \ref{ex_3} slightly by adding another component with additional free parameters $D$, $E$, and $F$,
\begin{equation}
f(x) = A\cos(B x+C) + D\cos(E x+F) \;\textrm{.}
\end{equation}
If the fit parameter $D$ becomes small such that $|D|\ll |A|$, the second component cannot influence the fit anymore and the two model parameters $E$ and $F$ are ``lost''. In simple words: This model may \emph{change} its flexibility during the fitting procedure.
\end{example}
Hence, for nonlinear models, $K$ may not even be constant.\footnote{For linear models this cannot happen, since products (or more complicated functions) of model parameters are nonlinear.} Of course, these two examples do \textit{not} verify the claim that always $K\neq N-P$ for nonlinear models. However, acting as counter-examples, they clearly falsify the claim that $K=N-P$ is always true for nonlinear models.

\subsection{Local linearisation}

As we have seen above, there is no well-defined method for estimating the number of degrees of freedom for a truly nonlinear model. We may now object that any well-behaved nonlinear model\footnote{With ``well-behaved'' we mean a model that can be differentiated twice with respect to all fit parameters, including mixed derivatives.} can be \textit{linearised} around the parameter values which minimise $\chi^2$. Let us denote the parameter values that minimise $\chi^2$ by $\vec\Theta$. We can then Taylor-expand $\chi^2$ at $\vec\Theta$ to second order, 
\begin{equation}\label{eq:CLT_expansion}
\chi^2(\vec\theta)\approx \chi^2_\textrm{min} + \sum_{p,q=1}^P \left.\frac{\partial^2\chi^2}{\partial\theta_p\partial\theta_q}\right|_{\vec\theta=\vec\Theta} (\theta_p-\Theta_p)(\theta_q-\Theta_q) \;\textrm{,}
\end{equation}
where the first derivative is zero at the minimum. Apparently, $\chi^2$ is now a quadratic function of the model parameters, i.e., the model is linearised. Does this mean that we can simply linearise the model in order to get rid of the problems with defining the number of degrees of freedom for a nonlinear model?

The answer is definitely no. The crucial problem is that linearisation is just an \textit{approximation}. The Taylor expansion of Eq. (\ref{eq:CLT_expansion}) has been truncated after the second-order term. There are two issues here: First, in general, we do not know how good this approximation really is for a given data sample. Second, we have no way of knowing how good the approximation \textit{needs} to be in order to sufficiently linearise the model from the number-of-degrees-of-freedom point of view.

Even if these issues did not concern us, would linearising the model really help? Again, the answer is no. As we have seen in Sect. \ref{sect:dofs_linear}, the number of degrees of freedom is also not necessarily given by $N-P$ for a linear model. Moreover, the truly worrisome result of Sect. \ref{sect:dofs_nonlinear} -- that the number of degrees of freedom is not constant -- is \textit{not} overcome by the linearisation. The reason is that the expansion of Eq.\ (\ref{eq:CLT_expansion}), and thereby the linearisation, depends nonlinearly upon \textit{where} the maximum is.\footnote{For a linear model the second derivatives of $\chi^2$ do not depend on any model parameters, i.e., they are constant.} Consequently, the uncertainties in the maximum position inherited from the data's noise propagate nonlinearly through the expansion of Eq.\ (\ref{eq:CLT_expansion}). Therefore, we have to draw the conclusion that there is no way of reliably estimating the number of degrees of freedom for a nonlinear model.

\subsection{Summary}

Summarising the arguments brought up so far, we have seen that estimating the number of degrees of freedom is absolutely nontrivial. In the case of linear models, the number of degrees of freedom is given by $N-P$ if and only if the basis functions are indeed linearly independent in the regime sampled by the given data. Usually, this is true in practice. Otherwise, the number of degrees of freedom is somewhere between $N-P$ and $N-1$. However, in the case of nonlinear models, the number of degrees of freedom can be anywhere between 0 and $N-1$ and it is even not necessarily constant during the fit. Linearising the model at the optimum does not really help to infer the number of degrees of freedom, because the linearised model still depends on the optimum parameters in a nonlinear way. Hence, it is questionable whether it is actually \textit{possible} to compute $\chi^2_\textrm{red}$ for nonlinear models in practice.

\section{Uncertainty in $\chi^2$\label{sect:noise_on_chi2}}

We now discuss another problem, which is completely independent of our previous considerations. Even if we were able to estimate the number of degrees of freedom reliably, this problem would still interfere with any inference based on $\chi^2_\textrm{red}$. This problem stems from the fact that the value of $\chi^2$ is subject to noise, which is inherited from the random noise of the data.\footnote{For a given set of data, $\chi^2$ can of course be computed. However, consider a second set of data, which was drawn from the same physical process such that only the noise realisation is different. For this second set of date, the value of $\chi^2$ will differ from that of the first set.} Consequently, there is an ``uncertainty'' on the value of $\chi^2$ and hence on $\chi^2_\textrm{red}$, which is typically ignored in practice. However, we show that this uncertainty is usually large and must not be neglected, because it may have a severe impact on the intended application.

Given some data with Gaussian noise, the true model having the true parameter values will generate a $\chi^2=N$ and has $N$ degrees of freedom because there is no fit involved. Hence, it results in a $\chi^2_\textrm{red}$ of 1. We therefore compare the $\chi^2_\textrm{red}$ of our trial model to 1 in order to assess convergence or to compare different models. Is this correct?

\begin{wrapfigure}{right}{8cm}
\includegraphics[width=8cm]{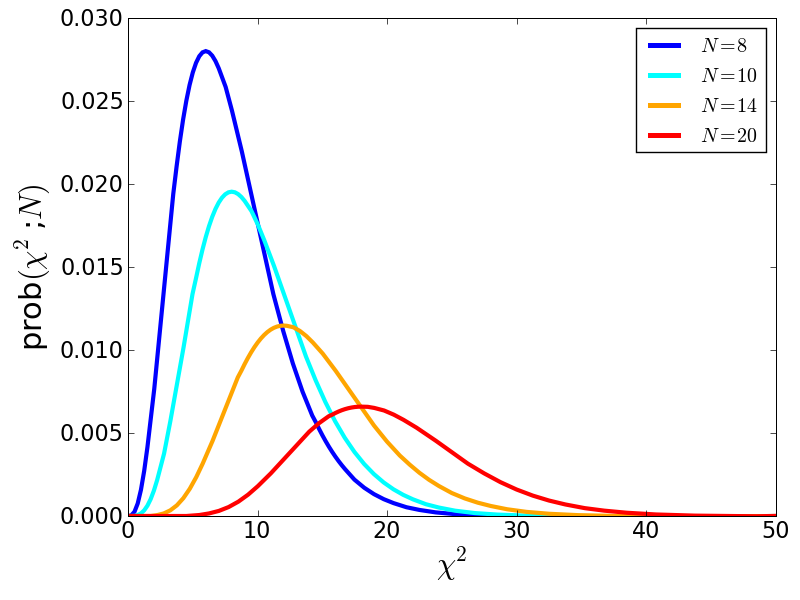}
\caption{$\chi^2$-distributions for different values of $K=N$ degrees of freedom. The distributions are asymmetric, i.e., mean and maximum (mode) do not coincide. For increasing $K=N$, the distributions become approximately Gaussian.}
\label{fig:chi2_distributions}
\end{wrapfigure}

In theory, yes. In practice, no. Even in the case of the true model having the true parameter values, where there is no fit at all, the value of $\chi^2$ is subject to noise. In this case, we are fortunate enough to be able to quantify this uncertainty. For the true model having the true parameter values and \textit{a-priori} known measurement errors $\sigma_n$, the normalised residuals,
\begin{equation}\label{eq:normalised_residuals}
R_n=\frac{y_n-f(\vec x_n,\vec\theta)}{\sigma_n}
\end{equation}
are distributed according to a Gaussian with mean $\mu=0$ and variance $\sigma^2=1$.\footnote{Again, we implicitely assume that the Gaussian errors are uncorrelated, as in Eq.\ (\ref{eq:def:chi2}). If the measurement errors $\sigma_n$ are not known a priori but have been estimated from the data, the normalised residuals $R_n$ are drawn from  Student's $t$-distribution \citep[e.g.][]{Barlow1993}. With increasing number of data points Student's $t$-distribution approaches a Gaussian distribution.} In this case only, $\chi^2$ is the sum of $K=N$ Gaussian variates and its probability distribution is given by the so-called $\chi^2$-distribution,
\begin{equation}\label{eq:chi2_distribution}
\prob(\chi^2;K) = \frac{1}{2^{K/2}\Gamma(K/2)} \left(\chi^2\right)^{K/2-1}e^{-\chi^2/2} \;\textrm{.}
\end{equation}
Figure \ref{fig:chi2_distributions} shows some $\chi^2$-distributions with different values of $K$. The expectation value of the $\chi^2$-distribution is,
\begin{equation}
\langle\chi^2\rangle = \int_0^\infty \chi^2\,\prob(\chi^2;K)\,d\chi^2 = K\;\textrm{.}
\end{equation}
In fact, this expectation value is sometimes used as an alternative definition of ``degrees of freedom''. As the $\chi^2$-distribution is of non-zero width, there is however an uncertainty on this expectation value. More precisely, the variance of the $\chi^2$-distribution is given by $2K$. This means the expectation value of $\chi^2_\textrm{red}$ for the true model having the true parameter values is indeed one, but it has a variance of $2/K=2/N$. If $N$ is large, the $\chi^2$-distribution becomes approximately Gaussian and we can take the root of the variance, $\sigma=\sqrt{2/N}$, as an estimate of the width of the (approximately Gaussian) peak. Let us consider a simple example in order to get a feeling how severe this problem actually is:
\begin{example}\label{ex_5}
We are given a data set comprised of $N=1,000$ samples. Let the task be to use $\chi^2_\textrm{red}$ in order to compare different models to select that one which fits the data best, or to fit a single model to this data and assess convergence. The true model having the true parameter values -- whether it is given or not -- will have a value of $\chi^2_\textrm{red}$ with an (approximated) Gaussian standard deviation of $\sigma=\sqrt{2/1000}\approx 0.045$. Consequently, within the $3\sigma$ interval $0.865\leq\chi^2_\textrm{red}\leq 1.135$ we can neither reliably differentiate between different models nor assess convergence.
\end{example}
This simple example clearly shows that this problem is very drastic in practice. Moreover, astronomical data sets are often much smaller than $N=1,000$, which increases the uncertainty of $\chi^2_\textrm{red}$.

Of course, there is not only an uncertainty on the comparison value of $\chi^2_\textrm{red}$ for the true model having the true parameter values. There is also an uncertainty on the value of $\chi^2_\textrm{red}$ for any other model. Unfortunately, we \textit{cannot} quantify this uncertainty via $\sigma\approx\sqrt{2/K}$ in practice anymore, because the $\chi^2$-distribution applies only to the true model having the true parameter values. For any other model the normalised residuals (cf.\ Eq.\ (\ref{eq:normalised_residuals})) are not Gaussian with mean $\mu=0$ and variance $\sigma^2=1$. Hence, $\chi^2$ is \textit{not} the sum of $K$ Gaussian variates and the derivation of the $\chi^2$-distribution is invalid.

\section{Alternative methods\label{sect:alternatives}}

If $\chi^2_\textrm{red}$ does not provide a reliable method for assessing and comparing model fits, convergence tests or error estimation, what other methods can then be used with more confidence? An in-depth survey of alternative methods would be beyond the scope of this manuscript. Therefore, we restrict our discussion on some outstanding methods. Concerning methods for error estimation, we refer the interested reader to \citet{Andrae2010d}.

\subsection{Residuals}

The first and foremost thing to do in order to assess the goodness of fit of some model to some data is to inspect the residuals. This is indeed trivial, because the residuals have already been computed in order to evaluate $\chi^2$ (cf. Eq. (\ref{eq:def:chi2})). For the true model having the true parameter values and \textit{a-priori} known measurement errors, the distribution of normalised residuals (cf. Eq. (\ref{eq:normalised_residuals})) is by definition Gaussian with mean $\mu=0$ and variance $\sigma^2=1$. For any other model, this is not true. Consequently, all one needs to do is to plot the distribution of normalised residuals in a histogram and compare it to a Gaussian of $\mu=0$ and $\sigma^2=1$. If the histogram exhibits a statistically \textit{significant} deviation from the Gaussian, we can rule out that the model is the truth. If there is no significant difference between histogram and Gaussian, this can mean (a) we found the truth, or (b) we do not have enough data points to discover the deviation. The comparison of the residuals to this Gaussian should be objectively quantified, e.g., by using a Kolmogorov-Smirnov test\footnote{The Kolmogorov-Smirnov (KS) test compares the empirical cumulative distribution function (CDF) of a sample to a theoretical CDF by quantifying the distance between the distributions. Under the (null) hypothesis that the sample \textit{is} from the given distribution, this distance (called the KS-statistic) has a known probability distribution. Now, the test calculates the KS-statistic and compares it to its known probability distribution.} \citep{Kolmogorov1933,Smirnov1948}.

In theory, this method may be used as a convergence diagnostic. In an iterative fit procedure, compare the distribution of normalised residuals to the Gaussian with $\mu=0$ and $\sigma^2=1$ in each iteration step, e.g., via a Kolmogorov-Smirnov test. At first, the likelihood of the residuals to be Gaussian will increase as the model fit becomes better. \textit{If} the fit finds a suitable local minimum, the model may eventually start to \textit{overfit} the data and the likelihood of the residuals to be Gaussian will decrease again, as the residuals will peak too sharply at zero. When this happens, the fitting procedure should be stopped. In practice, there is no guarantee that this works, as the fit may end up in a local minimum with residuals too widely spread to resemble the Gaussian with $\mu=0$ and $\sigma^2=1$.

\begin{wrapfigure}{right}{8cm}
\includegraphics[width=8cm]{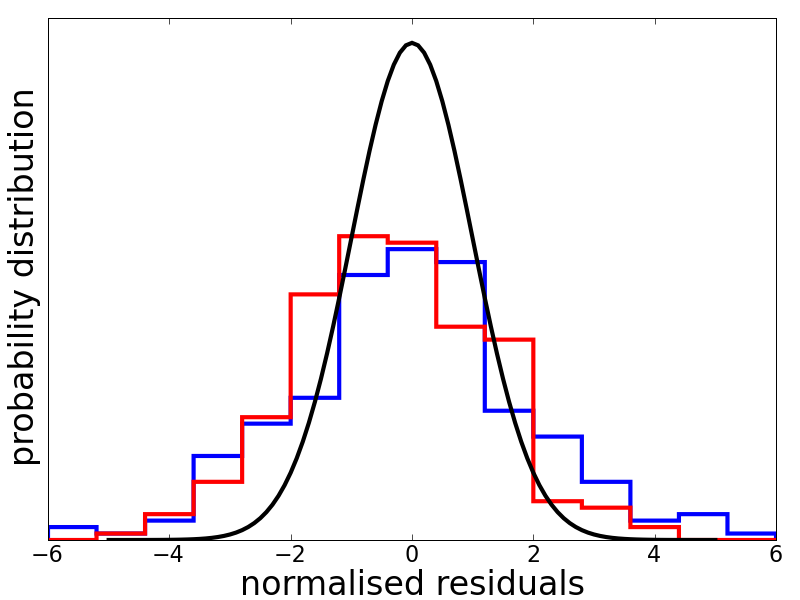}
\caption{Distributions of normalised residuals for the data and models of \citet{Vogt2010} using four planets (blue histogram) and six planets (red histogram). For comparison we also show the Gaussian with $\mu=0$ and $\sigma^2=1$. As the number of data points is large ($N=241$), there is no difference between this Gaussian and Student's $t$-distribution.}
\label{fig:dist_normed_residuals}
\end{wrapfigure}

Similarly, this method may also be used for model comparison. Given some data and a set of models, the model favoured by the data is that whose normalised residuals match the Gaussian with $\mu=0$ and $\sigma^2=1$ best. The winning model does not need to be the truth. Let us consider the following example:

\begin{example}\label{ex_6}
\citet{Vogt2010} analysed radial-velocity data of the nearby star GJ~581 and came to the conclusion that the data suggests the presence of six exoplanets instead of four as claimed by other authors using different data \citep[e.g.][]{Mayor2009}. \citet{Vogt2010} assumed circular orbits, which result in nonlinear models of the form of Eq.\ (\ref{eq:circular_orbit}) in Example \ref{ex_3}. Their claim that two additional planets are required is primarily justified from the associated $\chi^2_\textrm{red}$ (cf.\ Table \ref{table:KS_results}). We take the identical data as \citet{Vogt2010} and their asserted orbital parameters of the six planets (their Table 2). For every number of planets, we apply the KS-test to the normalised residuals. Table \ref{table:KS_results} gives the resulting $p$-values\footnote{Loosely speaking, in this case, the $p$-value is the probability that the true model can generate residuals that agree with the standard Gaussian as badly as or worse than the actually observed ones.} for the six models. In terms of the KS-test, the data used by \citet{Vogt2010} strongly favour the model using four planets over the model using six planets.  Furthermore, Fig.\ \ref{fig:dist_normed_residuals} displays the distributions of normalised residuals for the models using four and six planets. Evidently, both distributions deviate significantly from the Gaussian with $\mu=0$ and $\sigma^2=1$, which implies that neither model is compatible with the truth. The most likely explanation for this discrepancy is that planetary orbits may be elliptical, whereas \citet{Vogt2010} assumed circular orbits.
\end{example}

\begin{table}
\begin{tabular}{ccccccc}
\hline\hline
Planets & 1 & 2 & 3 & \textbf 4 & 5 & \textbf 6 \\
\hline
$p$-value & $8.71\cdot 10^{-11}$ & $2.97\cdot 10^{-9}$ & $2.51\cdot 10^{-7}$ & \boldmath$1.28 \cdot 10^{-4}$\unboldmath & $1.47\cdot 10^{-5}$ & $6.97\cdot 10^{-8}$ \\
$\chi^2_\textrm{red}$ & 8.426 & 4.931 & 4.207 & 3.463 & 2.991 & \textbf{2.506}
\end{tabular}
\caption{$p$-values from KS-test and $\chi^2_\textrm{red}$ for 1--6 planets for the data of \citet{Vogt2010} discussed in Example \ref{ex_6}.}
\label{table:KS_results}
\end{table}

There is a pitfall here, as well. The likelihood of the normalised residuals to come from a Gaussian with $\mu=0$ and $\sigma^2=1$ is also subject to noise, as in the case of the value of $\chi^2$. However, these uncertainties surely cannot explain the large differences in Table \ref{table:KS_results}.

\subsection{Cross-validation}

Cross-validation is one of the most powerful and most reliable methods for model comparison. Unfortunately, it is usually also computationally expensive. However, if $\chi^2_\textrm{red}$ is not applicable, e.g., because the model is nonlinear, computational cost cannot be used as an argument in disfavour of cross-validation.

The most straightforward (and also most expensive) flavour of cross-validation is ``leave-one-out cross-validation''. We are given $N$ data points and a set of models, and we want to know which model fits the data best. For each model, the goodness of fit is estimated in the following way:
\begin{enumerate}
\item Remove the $n$-th data point from the data sample.
\item Fit the model to the remaining $N-1$ data points.
\item Take the model fitted to the $N-1$ data points and compute its likelihood for the $n$-th data point that has been left out.
\item Repeat steps 1 to 3 from $n=1$ to $n=N$ and compute the goodness of the prediction of the whole data set by multiplying the likelihoods obtained in step 3.
\end{enumerate}
Steps 3 and 4 require the data's error distribution to be known in order to evaluate the goodness of the prediction for the left-out data point through its likelihood. For instance, if the data's errors are Gaussian, the goodness of the prediction is simply given by Eq. (\ref{eq:normalised_residuals}) as usual. Evidently, repeating steps 1 to 3 $N$ times is what makes leave-one-out cross-validation computationally expensive. It is also possible to leave out more than one data point in each step.\footnote{The reason why cross-validation is so reliable is that it draws on the predictive error of the model, rather than the fitting error. Therefore, cross-validation can detect underfitting (the model is not flexible enough to describe the data) and also overfitting (the model is too flexible compared to the data). The fitting error is only sensitive to underfitting, but not to overfitting ($\chi^2$ always decreases if the model becomes more complex).} However, if the given data set is very small, cross-validation becomes unstable. A nice application of cross-validation can be found, e.g., in \citet{Hogg2008}.

\subsection{Bootstrapping}

Bootstrapping is somewhat more general than cross-validation, meaning it requires less knowledge about the origin of the data. Cross-validation requires the data's error distribution to be known in order to evaluate the likelihoods, whereas bootstrapping does not. Of course, this is an advantage if we do \textit{not} have this knowledge. However, if we do know the data's errors, we should definitely exploit this knowledge by using cross-validation. Bootstrapping is discussed in \citet{Andrae2010d} in the context of error estimation. Therefore, we restrict its discussion here on the context of model comparison.

Let us suppose we are given 4 measurements $y_1,y_2,y_3,y_4$. We then draw subsamples of size 4 from this data set with replacement. These subsamples are called bootstrap samples. Examples are:
\begin{itemize}
\item $y_1,y_2,y_3,y_4$ itself,
\item $y_1,y_2,y_2,y_4$,
\item $y_2,y_4,y_4,y_4$,
\item $y_1,y_1,y_1,y_1$,
\item etc.
\end{itemize}
We draw a certain number of such bootstrap samples, and to every such sample we then fit all the models that are to be compared.

In the context of model comparison, bootstrapping is typically used as ``leave-one-out bootstrap'' \citep[e.g.][]{Hastie2009}. The algorithm is given by:
\begin{enumerate}
\item Draw a certain number of bootstrap samples from a given data set.
\item Fit all the models to every bootstrap sample.
\item For the $n$-th data point $y_n$ in the given data set, consider only those bootstrap samples that do \textit{not} contain $y_n$. Predict $y_n$ from the models fitted to these bootstrap samples.
\item Repeat step 3 from $n=1$ to $n=N$ and monitor the goodness of the predictions, e.g., by least squares.
\end{enumerate}
Like cross-validation, bootstrapping aims at the prediction error of the model. Therefore, it is sensitive to over- and underfittings.

\section{Conclusions\label{sect:conclusions}}

We have argued that there are two fundamental problems in using $\chi^2_\textrm{red}$, which are completely independent of each other:
\begin{enumerate}
\item In Sect. \ref{sect:dofs}, we have seen that estimating the number of degrees of freedom, which is necessary for evaluating $\chi^2_\textrm{red}$, is absolutely nontrivial in practice:
\begin{itemize}
\item Concerning \textit{linear} models, for $N$ given data points and $P$ fit parameters the number of degrees of freedom is \textit{somewhere} between $N-P$ and $N-1$, where it is $N-P$ if and only if the basis functions of the linear model are linearly independent for the given data. Equation (\ref{eq:P_effective}) provides a quantification for the effective number of fit parameters of a linear model. Priors can cause a linear model to become nonlinear.
\item Concerning \textit{nonlinear} models, the number of degrees of freedom is \textit{somewhere} between zero and $N-1$ and it may not even be constant during a fit, i.e., $N-P$ is a completely unjustified guess. The authors are not aware of any method that reliably estimates the number of degrees of freedom for nonlinear models. Consequently, it appears to be impossible to compute $\chi^2_\textrm{red}$ in this case.
\end{itemize}
\item In Sect. \ref{sect:noise_on_chi2}, we have seen that the actual value of $\chi^2_\textrm{red}$ is \textit{uncertain}. If the number $N$ of given data points is large, the uncertainty of $\chi^2_\textrm{red}$ is approximately given by the Gaussian error $\sigma=\sqrt{2/N}$. For $N=1,000$ data points, this means that within the $3\sigma$-interval $0.865\leq\chi^2_\textrm{red}\leq 1.135$ we cannot compare models or assess convergence.
\end{enumerate}
Given these considerations, it appears highly questionable whether the popularity of $\chi^2_\textrm{red}$ -- which is certainly due to its apparent simplicity -- is indeed justified. As a matter of fact, $\chi^2_\textrm{red}$ cannot be evaluated for a nonlinear model, because the number of degrees of freedom is \textit{unknown} in this case. This is a severe restriction, because many relevant models are nonlinear. Moreover, even for linear models, $\chi^2_\textrm{red}$ has to be used with due caution, considering the uncertainty in its value.

Concerning alternative methods for model comparison, we have explained cross-validation and bootstrapping in Sect.\ \ref{sect:alternatives}. We also explained how the normalised residuals of a model can be used to infer how close this model is to the true model underlying the given data. Concerning alternative methods for error estimation, we refer the interested reader to \citet{Andrae2010d}.

Finally, we want to emphasise that the above considerations concerning $\chi^2_\textrm{red}$ have \textit{no} impact on the correctness of minimising a $\chi^2$ in order to fit a model to data. Fitting models to data is a completely different task that should not be confused with model comparison or convergence testing. Minimising $\chi^2$ is the correct thing to do whenever the data's measurement errors are Gaussian and a maximum-likelihood estimate is desired.

\paragraph*{Acknowledgements} RA thanks David Hogg for detailed discussions on this subject. David Hogg also came up with a couple of the examples mentioned here. Furthermore, RA thanks Coryn Bailer-Jones for helpful comments on the contents of this manuscript. RA is funded by a Klaus-Tschira scholarship. TS is funded by a grant from the Max Planck Society. PM is supported by the DFG Priority Programme 1177.

\bibliographystyle{aa}

\def\physrep{Phys. Rep.}%
          % Astrophysical Journal, Supplement
\def\apjs{ApJS}%
          % Astrophysical Journal, Supplement
\def\apj{ApJ}%
          % Astrophysical Journal
\def\apjl{ApJL}%
          % Astrophysical Journal
\def\aj{AJ}%
          % Astronomical Journal
\def\aap{A\&A}%
          % Astronomy and Astrophysics
\def\aaps{A\&AS}%
          % Astronomy and Astrophysics Supplements
\def\mnras{MNRAS}%
          % Monthly Notices of the RAS
        
\bibliography{bibliography}

\end{document}